\documentclass[%
 reprint,
 amsmath,amssymb,
prl,
]{revtex4-2}

\usepackage{graphicx}
\usepackage{dcolumn}
\usepackage{bm}

\usepackage{color}

\begin{document}

\preprint{APS/123-QED}

\title{Characterizing Data Assimilation in Navier--Stokes Turbulence\\
with Transverse Lyapunov Exponents}

\author{Masanobu Inubushi$^{1,2}$}%
\email[]{inubushi@rs.tus.ac.jp}

\author{Yoshitaka Saiki$^{3}$}

\author{Miki U. Kobayashi$^{4}$}

\author{Susumu Goto$^{2}$}%
\affiliation{$^{1}$Department of Applied Mathematics, Tokyo University of Science, Tokyo 162-8601, Japan\\
$^{2}$Graduate School of Engineering Science, Osaka University, Osaka 560-8531, Japan\\
$^{3}$Graduate School of Business Administration, Hitotsubashi University, Tokyo 186-8601, Japan\\
$^{4}$Faculty of Economics, Rissho University, Tokyo 141-8602, Japan
}

\date{\today}

\begin{abstract}
Data assimilation (DA) reconstructing small-scale turbulent structures is crucial for forecasting and understanding turbulence.
This study proposes a theoretical framework for DA based on ideas from chaos synchronization, in particular, the transverse Lyapunov exponents (TLEs).
The analysis with TLEs characterizes a critical length scale, below which the turbulent dynamics is synchronized to the larger-scale turbulent dynamics, indicating successful DA.
An underlying link between TLEs and the maximal Lyapunov exponent suggests that the critical length scale depends on the Reynolds number. 
Furthermore, we discuss new directions of DA algorithms based on the proposed framework.
\end{abstract}

\maketitle


\begin{figure*}
\includegraphics[width=150mm]{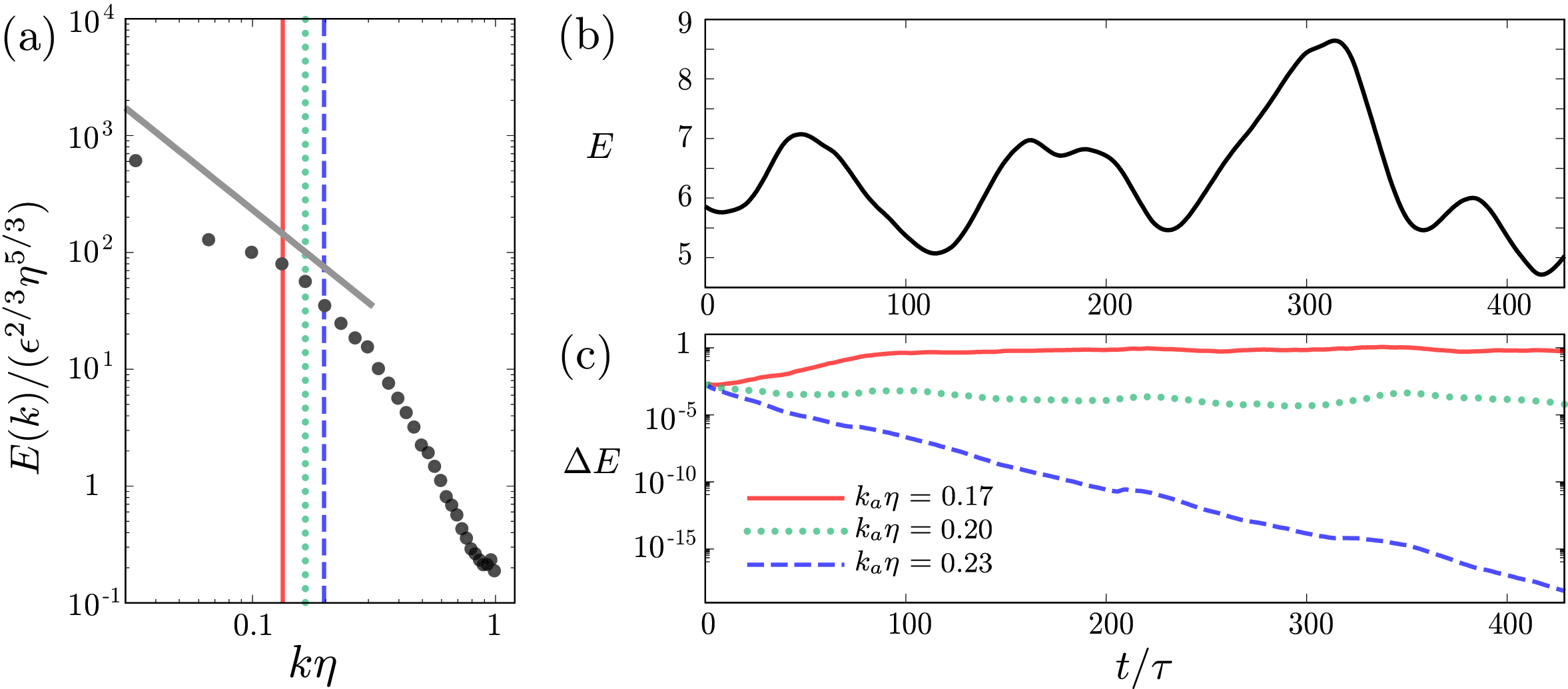}
\caption{Continuous data assimilation in the box turbulence. (a) Energy spectrum $E(k)$ and (b) time series of kinetic energy $E(t)$ in statistically steady states,
where the energy dissipation rate $\epsilon$, Kolmogorov time $\tau$, and length $\eta$ are used for the normalization.
The gray line in (a) represents a slope of $k^{-5/3}$.
(c) Time series of energy of difference fields between ${\bm u}_{1}(t)$ and ${\bm u}_{2}(t)$
with three different $k_{a}$:
$k_{a}\eta=0.17$ (solid red), $0.20$ (dot green), and $0.23$ (dashed blue).
As a reference, the vertical lines corresponding to each $k_{a}$ are depicted in (a).
}
\label{fig1}
\end{figure*}

Predicting the future states of Navier--Stokes turbulence is 
a huge challenge due to its chaotic dynamics over broad spatiotemporal scales.
In particular, observational data on small-scale turbulent structures are generally unavailable; therefore, there is uncertainty in the initial conditions for prediction.
These small-scale uncertainties increase exponentially fast,
considering the Lyapunov exponent $\lambda$ of three-dimensional turbulence
is mainly determined by the Kolmogorov time scale $\tau$~\cite{Ruelle79} as $\lambda \propto 1/\tau$.
This exponential error growth from the small scales finally limits the predictability of large-scale motions~\cite{Lorenz, Leith-Kraichnan}.
Therefore, for predicting turbulence, it is crucial to infer small-scale turbulent structures from only observational data of large-scale ones.

Data assimilation (DA) is suitable for such inferences.
Previous studies have shown critical length scales below which turbulent structures can be inferred via DA methods with only observational data of larger-scale structures~\cite{Titi, YYK2005,Lalescu2013, Martin2021, Zaki, Biferale, Li2020}, i.e.,
the small-scale turbulent dynamics are reconstructed or `slaved' by the larger-scale dynamics.
Interestingly, in three-dimensional turbulence,
a common critical length scale,
approximately $20 \eta$ where $\eta$ is the Kolmogorov length,
has been reported irrespective of the details of the DA algorithms.
This length scale corresponds to the wavenumber $k^{\ast}=0.2/ \eta$,
which was first found in the continuous DA~\cite{YYK2005}
and recently found in four-dimensional variational DA~\cite{Li2020} and nudging method~\cite{Biferale}.
This indicates that the slaving small-scale dynamics can be understood based on the nature of the Navier--Stokes equations rather than a specific DA algorithm, which is essential for turbulence physics~\cite{GSK2017, YGT2022} and modeling~\cite{MIG}, as discussed later.
However, the physical origins of the critical length scale and the Reynolds-number dependence remain unclear.

In this Letter, we propose a theoretical framework for studying such DA phenomena as a {\it stability problem}.
The proposed framework explains, for the first time, how the critical length scale can be determined from the property of the Navier--Stokes equations.
Inspired by the concept of Blowout bifurcation~\cite{Ott} in the study of chaos synchronization, we introduce an invariant manifold,
{\it DA manifold}, in phase space and present a stability analysis,
wherein the {\it transverse} Lyapunov exponents (TLEs) characterize the critical length scale, determining the success or failure of the DA process.

Moreover, we show that the TLEs are a generalization of the maximal Lyapunov exponent $\lambda$ of the turbulence attractor,
whose Reynolds number dependence has been extensively studied in research on unpredictability~\cite{Ruelle79, Boffetta2017, Arjun, Mohan}.
Considering this relationship between the TLEs and maximal Lyapunov exponent and their Reynolds number dependency, we conclude that the critical length scale, $k^{\ast}\eta$, depends on the Reynolds number.
The findings of this study suggest novel directions for practical DA research based on stability, and moreover, 
shed light on the fundamental relationship between the small-scale dynamics slaved to the larger-scale dynamics
and the unpredictability of turbulence.

{\it Formulation---}
On the basis of the mathematical analysis by Olson and Titi (2003)~\cite{Titi},
we study a twin experiment of the continuous DA defined by two incompressible Navier--Stokes equations ($i=1,2$):
\begin{align}
\partial_{t}{\bm u}_{i} +({\bm u}_{i} \cdot \nabla){\bm u}_{i} &= - \nabla \pi_{i} + \nu \Delta {\bm u}_{i} + {\bm f},\nonumber\\
\nabla \cdot {\bm u}_{i} &=0, \label{nse}
\end{align}
on the $d$-dimensional periodic domain $[0, L]^{d}$,
where $\nu$ denotes the kinematic viscosity.
The first vector field ${\bm u}_{1}({\bm x},t)$ is the {\it true} velocity field, which is used as a reference;
and the second, ${\bm u}_{2}({\bm x},t)$, is used for the DA process (the twin system).
Here, $\pi_{i}~(i=1,2)$ is the pressure of each system,
and ${\bm f}$ is the external forcing.
The projection operators $P_{k_{a}}$ and $Q_{k_{a}}$ are introduced to the Fourier representation as follows:
\begin{align}
P_{k_{a}} {\bm u} = \sum_{| {\bm k} | < k_{a}} \widehat{\bm u}_{\bm k} e^{i {\bm k} \cdot {\bm x}},~~Q_{k_{a}} = I - P_{k_{a}}, \label{PQ}
\end{align}
where $\widehat{\bm u}_{\bm k}$ is the Fourier coefficient of the velocity field $\bm u$ corresponding to the wavenumber vector
${\bm k} \in \{ 2\pi {\bm m} / L : {\bm m} \in \mathbb{Z}^{d} \}$.
For each wavenumber $k_{a} \in \mathbb{R}$,
which is a key control parameter in this study,
these operators decompose the velocity fields into large-scale ${\bm p}_{i}=P_{k_{a}} {\bm u}_{i}$ and small-scale ${\bm q}_{i}=Q_{k_{a}} {\bm u}_{i}$ parts, i.e., ${\bm u}_{i} = {\bm p}_{i} + {\bm q}_{i}~~(i=1,2)$.

The continuous DA method assumes that the large-scale structure of the true velocity field ${\bm p}_{1}(t)$ can be observed at all times $t\ge 0$ without observational errors.
Therefore, partial observational data ${\bm p}_{2}(t)$ are used for the twin system as: ${\bm p}_{2}(t) \equiv {\bm p}_{1} (t)$.
Hence, ${\bm u}_{2}(t)={\bm p}_{1}(t)+ {\bm q}_{2}(t)$.
The evolution equations for ${\bm q}_{2}(t)$ are derived from Eq.~(\ref{nse}) for $i=2$ using $Q_{k_{a}}$:
\begin{align}
\partial_{t} {\bm q}_{2} + Q_{k_{a}} ({\bm u}_{2} \cdot \nabla {\bm u}_{2}) &= - \nabla \pi'_{2} + \nu \Delta {\bm q}_{2} + Q_{k_{a}} {\bm f},\nonumber\\
\nabla \cdot {\bm q}_{2} &= 0, \label{da}
\end{align}
where $\pi'_{2} :=Q_{k_{a}} \pi_{2} $~\cite{Titi}.
The goal of the continuous DA method is to infer the small-scale structure of the true velocity field ${\bm q}_{1}(t)$ using Eq.~(\ref{da}) with the observational data of ${\bm p}_{1}(t)$.
In the two-dimensional case ($d=2$), Olson and Titi (2003) rigorously showed a sufficient condition for successful DA;
for a given kinematic viscosity $\nu$ and forcing term ${\bm f}$,
there exists a critical wavenumber $k^{\ast}_{a}$ such that, if $k_{a} > k^{\ast}_{a}$, then
${\bm q}_{2}(t)$ converges to ${\bm q}_{1}(t)$ exponentially, i.e., ${\bm u}_{2}(t) \to {\bm u}_{1}(t)~(t \to \infty)$.
Therefore, continuous DA enables us to infer the small-scale structure of the true velocity field ${\bm q}_{1}$
without direct observation.

In the three-dimensional case ($d=3$), 
Yoshida, Yamaguchi, and Kaneda (2005)~\cite{YYK2005} studied continuous DA using direct numerical simulations of the Navier--Stokes equations with assimilation at each time step.
Starting with the initial velocity fields of ${\bm u}_{1}(t)$ and ${\bm u}_{2}(t)$,
the velocity fields ${\bm u}_{1}(t+ \Delta t)$ and ${\bm u}_{2}(t + \Delta t)$ were calculated independently
using the fourth-order Runge--Kutta method
and ${\bm p}_{2} (t + \Delta t)$ was replaced by the true state ${\bm p}_{1} (t + \Delta t)$.
Thus, $\tilde{\bm u}_{2}(t + \Delta t):={\bm p}_{1} (t + \Delta t) + {\bm q}_{2} (t + \Delta t)$
where $\tilde{\cdot}$ denotes the updated state.
Then, the time evolution was calculated using the initial conditions
${\bm u}_{1}(t+ \Delta t)$ and $\tilde{\bm u}_{2}(t+ \Delta t)$ independently.
Using this method, the critical wavenumber $k_{a}^{\ast}$ was identified as $k_{a}^{\ast} \eta = 0.2$, where $\eta$ is the Kolmogorov length~\cite{YYK2005}.

{\it Numerical experiments---}
We conducted direct numerical simulations of the three-dimensional Navier--Stokes equations in a periodic box with $L=2 \pi$
driven by a steady forcing
${\bm f}(x,y,z) =(
-\sin x \cos y,
\cos x \sin y,
0)^T$.
Figs.~1~(a) and (b) show, respectively, the energy spectrum of the turbulence
and the time series of kinetic energy $E$ in a statistically steady state with Reynolds number Re $=570$.
More details of the setup for the numerical experiments can be found in the Supplemental Material.

\begin{figure*}
\includegraphics[width=160mm]{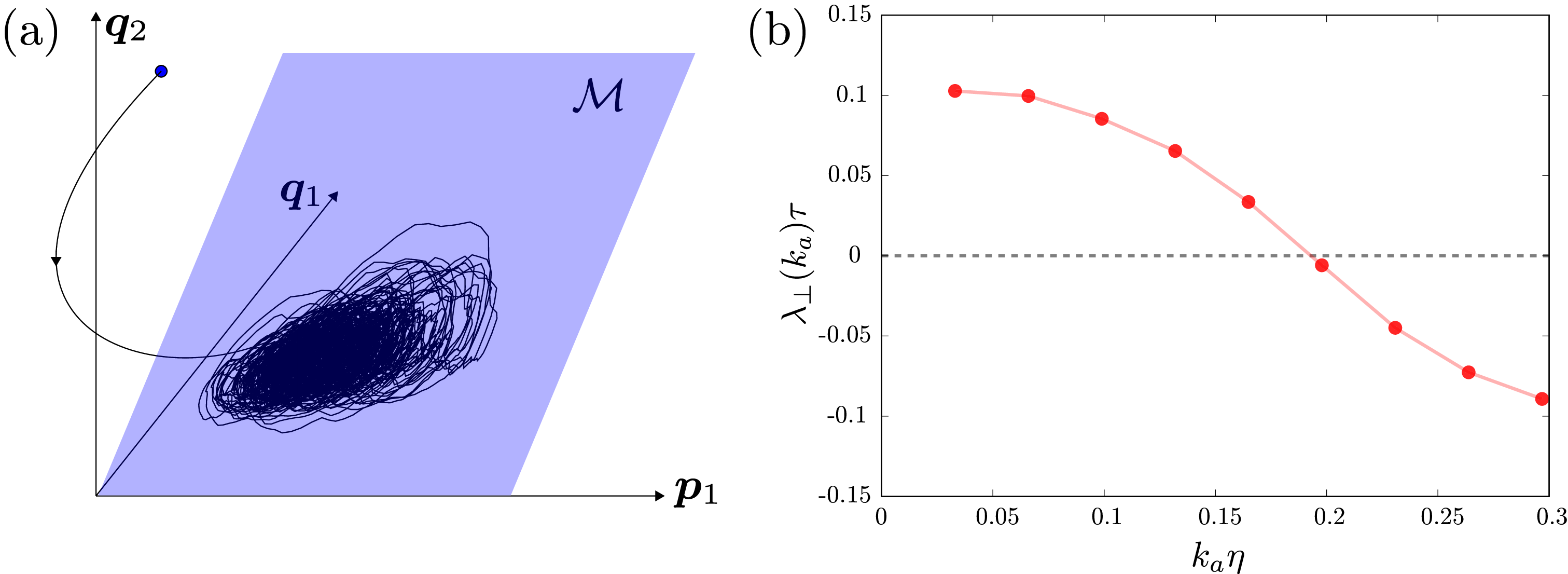}
\caption{(a) Schematics of phase space and the DA manifold $\mathcal{M}$,
illustrating the successful DA process, where the solution trajectory from the initial point (the blue dot) approaches $\mathcal{M}$.
(b) Transverse Lyapunov exponents (TLEs) $\lambda_{\perp}(k_{a})$ as a function of the wavenumber $k_{a}$ normalized by the Kolmogorov time $\tau$ and length $\eta$. The sign changes from positive to negative at $k_{a} \eta \simeq 0.2$, characterizing the success or failure of the DA process shown in Fig.~\ref{fig1} (c).}
\label{fig2}
\end{figure*}

The initial condition of the true system is a turbulent field in a statistically steady state ${\bm u}_{1}(0) ={\bm p}_{1} (0) + {\bm q}_{1} (0)$.
We obtained the initial condition of the twin system ${\bm u}_{2}(0)$
by adding a perturbation only to ${\bm q}_{1} (0)$.
This procedure is similar to that adopted by Yoshida, Yamaguchi, and Kaneda (2005)~\cite{YYK2005}.
Fig.~1~(c) shows the time series of energy of the difference field between ${\bm u}_{1}(t)$ and ${\bm u}_{2}(t)$,
$\Delta E(t)= 1/2| {\bm u}_{1} - {\bm u}_{2}|^{2}$, where $|{\bm u}|^{2} = \sum_{k} | \widehat{\bm u}_{k}|^{2}$,
for three values of $k_{a}$:
$k_{a}\eta=0.17$ (solid red), $0.20$ (dotted green), and $0.23$ (dashed blue).
Although the energy of the difference field does not decrease for $k_{a}\eta=0.17$ and $0.20$,
it decreases exponentially for $k_{a}\eta =0.23$, thereby indicating a successful DA process.
In other words, when $k_{a}\eta =0.23$,
small-scale structures of the velocity field can be determined by the sequential data of large-scale structures.
This result is quantitatively the same as that of the previous studies~\cite{YYK2005,Lalescu2013,Li2020,Biferale};
i.e., irrespective of differences in the forcing terms and details of the DA methods,
the critical wavenumber is $k_{a}^{\ast}\eta = 0.2$.
For the spatiotemporal dynamics of vortex structures reconstructed using the DA process, see the movie in the Supplemental Material.

{\it DA manifold and its stability---}
We characterize the critical wavenumber with a stability property of the (skew-product) dynamical system determined by
the Navier--Stokes equations,
which can be expressed as
$\frac{\partial}{\partial t} [ {\bm p}_{1}, {\bm q}_{1} ]^{T} = {\bm F}({\bm p}_{1}, {\bm q}_{1}) $ (the base system),
and Eq.~(\ref{da}),
$\frac{\partial}{\partial t}  {\bm q}_{2}  = {\bm G}({\bm p}_{1}, {\bm q}_{2}) $ (the fiber system).
We focus on the manifold defined by $\mathcal{M}=\{ ({\bm p}_{1}, {\bm q}_{1}, {\bm q}_{2}) ~|~ {\bm q}_{1}= {\bm q}_{2} \}$,
which is invariant because the solution trajectory starting from an initial point on $\mathcal{M}$ stays on there.
We refer to $\mathcal{M}$ as a {\it DA manifold}. Fig.~\ref{fig2}~(a) shows schematics of the solution trajectory and $\mathcal{M}$ in the phase space.

The success of the continuous DA process implies asymptotic stability of $\mathcal{M}$.
Let us now consider a successful DA process, i.e., the solution trajectory starting an initial point apart from $\mathcal{M}$, i.e., ${\bm q}_{1}(0) \neq {\bm q}_{2}(0)$,
converges to $\mathcal{M}$ asymptotically in time; that is, ${\bm q}_{2}(t) \to {\bm q}_{1} (t)~(t \to + \infty)$.
This can be interpreted as $\mathcal{M}$ being asymptotically stable.
The linear stability analysis of $\mathcal{M}$ gives a priori knowledge on whether the DA process succeeds or fails.
To this end, we introduce an infinitesimal perturbation to the velocity field $\delta {\bm q}={\bm q}_{2}-{\bm q}_{1}$
and derive the variational equations as follows:
\begin{equation}
  \begin{split}
\partial_{t} \delta {\bm q}  + Q_{k_{a}} ( {\bm u}_{1} \cdot \nabla \delta {\bm q} )+Q_{k_{a}} ( \delta {\bm q} \cdot \nabla {\bm u}_{1} ) \\
= - \nabla \delta \pi + \nu \Delta  \delta {\bm q}, \label{ve}
  \end{split}
\end{equation}
where $\delta \pi = \pi'_{2} - \pi'_{1}$ is the perturbation in the pressure field
(see for the derivation in Supplemental Material).
The transverse Lyapunov exponent (TLE) is defined as
\begin{equation}
\lambda_{\perp} (k_{a}):= \lim_{t \to \infty} \frac{1}{t} \ln | \delta {\bm q}(t) |,
\end{equation}
if the limit exists.
A negative TLE, $\lambda_{\perp}<0$, indicates asymptotic linear stability of the DA manifold $\mathcal{M}$,
which implies a successful DA process.
By contrast,
if the TLE is positive, $\lambda_{\perp}>0$, the DA manifold $\mathcal{M}$ is linearly unstable, which implies a failure of the DA process.
The TLE characterizes the average exponential growth or decay rate of the norm of the perturbation along the solution trajectory within $\mathcal{M}$.

The TLEs $\lambda_{\perp} (k_{a})$ explain the results of the numerical experiments for continuous DA in the Navier--Stokes turbulence, as shown in Fig.~1 (c).
The numerical integration of the variational equations (\ref{ve}) coupled with the Navier--Stokes equations (\ref{nse}) for ${\bm u}_{1}$ gives a TLE $\lambda_{\perp}(k_{a})$ for each fixed $k_{a}$.
Fig.~\ref{fig2} (b) shows the normalized TLE $\lambda_{\perp}(k_{a})\tau$ as a function of the normalized wavenumber, $k_{a}\eta$.
For $k_{a}\eta<0.2$, the TLEs are positive, $\lambda_{\perp}>0$; that is, the DA manifold $\mathcal{M}$ is unstable.
The TLE decreases as $k_{a}$ increases and becomes negative for $k_{a}\eta>0.2$; that is, $\mathcal{M}$ is stable.
The change in stability of $\mathcal{M}$ at the critical wavenumber $k_{a}^{\ast} \simeq 0.2/\eta$ explains the results of the success or failure of the DA process shown in previous studies~\cite{YYK2005,Lalescu2013, Biferale} and Fig.~1~(c).

\begin{figure}
\includegraphics[width=\linewidth]{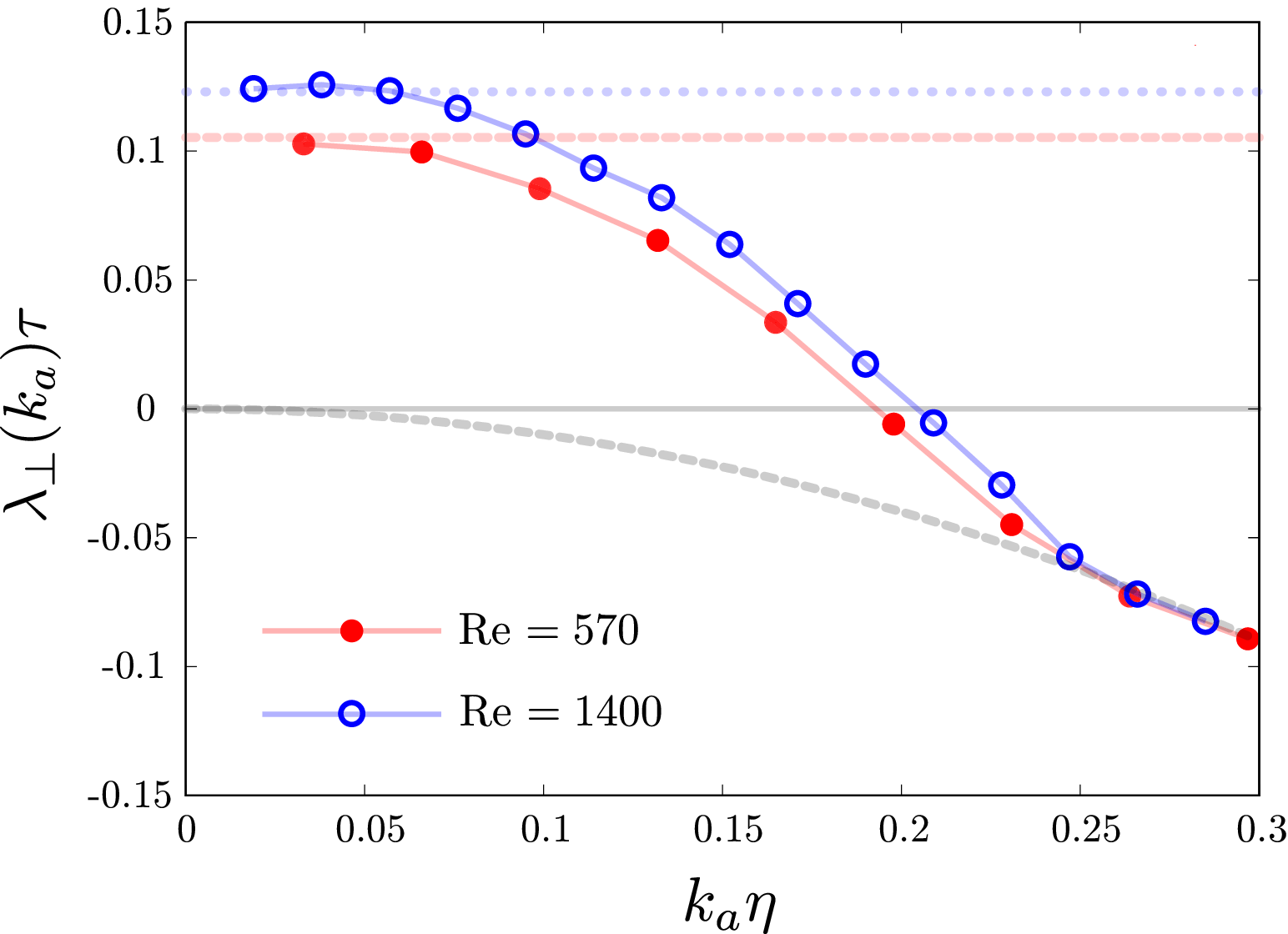}
\caption{Reynolds-number dependence of TLEs.
The normalization is the same in Fig.~\ref{fig2} (b).
The red solid and blue open circles represent the normalized TLEs for Re $= 570$ and Re $= 1400$, respectively.
The horizontal red dashed and blue dotted lines indicate the values of the normalized Lyapunov exponents $\lambda \tau$ for Re $= 570$ and Re $= 1400$, respectively.
The gray dashed curve shows $\lambda_{\perp}(k_{a}) \tau = - (k_{a} \eta)^{2}$, where the viscous term determines the perturbation dynamics.
}
\label{fig3}
\end{figure}

{\it Reynolds-number dependence---}
To study the Reynolds number dependence, the normalized TLEs $\lambda_{\perp} \tau$ for Re $= 1400$ are denoted as blue open circles in Fig.~\ref{fig3}.
For reference, the red circles denote the TLEs $\lambda_{\perp} (k_{a}) \tau$ for Re $= 570$, which are the same as those in Fig.~\ref{fig2} (b).
The critical wavenumber $k_{a}^{\ast}\eta$ defined by the sign change of the TLEs {\it weakly} depends on the Reynolds number;
for Re $= 1400$ $k_{a}^{\ast}\eta$ shifts to a larger value than Re $= 570$.
To understand this weak Re dependence of the critical wavenumber, we consider the asymptotic forms of the TLE $\lambda_{\perp}(k_{a})$ at the small and large
wavenumber $k_{a}$, respectively.

In the small wavenumber limit,
the TLE is reduced to the maximal Lyapunov exponent of the turbulent attractor.
The variational equations (\ref{ve}) describe the perturbation dynamics confined to the wavenumber regions higher than $k_{a}$.
As $k_{a}$ decreases, the perturbation dynamics become less confined.
At $k_{a}=0$, the perturbation can evolve in the tangent space in any direction; that is, no confinement.
In this case, $Q_{k_{a}}=I$ (the identity operator), and the variational equations (\ref{ve}) are reduced to the variational equations of the Navier--Stokes equations,
under which the TLE reduces to the maximal Lyapunov exponent of the turbulent attractor, that is $\lambda_{\perp}(0)=\lambda$.

The horizontal red dashed and blue dotted lines in Fig.~\ref{fig3} show the values of the normalized Lyapunov exponents $\lambda \tau$ for Re $= 570$ and Re $= 1400$, respectively.
The TLEs for each Re number converge to the normalized Lyapunov exponents as $k_{a} \to +0$.
For Re $= 570$, the value of the maximal Lyapunov exponent is $\lambda \tau \simeq 0.1$, and it increases with Re.
These results agree with the recent findings~\cite{Boffetta2017, Mohan, Arjun} claiming that the maximal Lyapunov exponent
increases with Re faster than predicted by dimensional analysis, that is, $\lambda \propto 1/\tau$.
In particular, the lower inset of Fig.~4 of Boffetta and Musacchio (2017)~\cite{Boffetta2017} shows that $\lambda \tau$ is an increasing function of the {\it logarithm} of Re.
Therefore, $\lambda \tau$  depends on the Reynolds number, although this dependence is weak.

Second, for the large-wavenumber limit of $\lambda_{\perp}(k_{a})$, the perturbation is confined to the higher-wavenumber region
where the viscous term is dominant and $\partial_{t} \delta {\bm q} \sim \nu \Delta \delta {\bm q}$.
This suggests that $\lambda_{\perp}(k_{a}) \tau = - (k_{a} \eta)^{2}$, as denoted by the gray dashed curve in Fig.~\ref{fig3}.
The TLEs for different Reynolds numbers collapse onto the curve for $k_{a} \eta \gtrsim 0.25$.

In summary, the TLEs $\lambda_{\perp}(k_{a})$ shown in Fig.~\ref{fig3} connect the maximal Lyapunov exponents at $k_{a}=0$
and the curve $ - (k_{a} \eta)^{2}$ for the large $k_{a}$.
In addition, the maximal Lyapunov exponent,
$\lambda_{\perp}(0)=\lambda$, increases slightly with Re~\cite{Boffetta2017,Mohan, Arjun}.
These findings indicate that
as the Reynolds number increases,
there is an upward shift of $\lambda_{\perp}(k_{a})\tau$
and a slight increase in the critical wavenumber $k_{a}^{\ast}$.

{\it Discussion and Conclusion---}
DA is becoming an increasingly significant tool in data-driven forecasting.
However, little is known about the critical wavenumber $k^{\ast}$, which plays a central role in various DA methods for three-dimensional turbulence~\cite{YYK2005, Biferale, Lalescu2013, Martin2021, Zaki}.
This study establishes a novel framework based on the theories of stability and Blowout bifurcation~\cite{Ott} and clarifies the critical wavenumber not from the results of DA but from the TLEs, which are the characteristic quantities of the Navier--Stokes equations.
Furthermore,
considering the novel discovery of the Reynolds number dependence of the maximal Lyapunov exponent~\cite{Boffetta2017, Mohan, Arjun},
the relationship between the TLEs and the maximal Lyapunov exponent suggests a weak Reynolds number dependence of the critical wavenumber for the first time.

This Letter aims to present novel concepts completely different from those used in the well-established DA research fields~\cite{Titi, Sebastian};
thus, a systematic investigation of the Reynolds number dependence of TLEs,
in particular, the critical length scale,
is beyond the scope of the present study and a crucial future challenge.
To this end, developing efficient algorithms for calculating the TLEs would be helpful.
The continuous DA is an ideal setting for the first step; extensions of our framework to incorporate the presence of noise and mismatch of the Reynolds number
will be important not only in practice
but also in the research of high-dimensional chaos synchronization.

Besides phase-space dynamics studied in this Letter,
understanding the turbulent dynamics in {\it physical} space will be complementarily necessary.
Remarkably,
the critical wavenumber, $k^{\ast}=0.2/ \eta$, has been identified in a context that differs from DA; that is,
a recent study on vortex stretching found the far dissipation range as the wavenumber region above $k^{\ast}$, i.e., $k>k^{\ast}$ \cite{YGT2022}.
In terms of the Kolmogorov--Richardson energy cascade,
the turbulent dynamics in the far dissipation range terminates the cascade process.
Although structures in the range acquire the energy from larger scales, they cannot transfer it to smaller ones but dissipate it there instead.
This may imply that they are slaving to larger-scale structures and gives an interpretation of the small-scale slaving dynamics in the DA context.

In addition to this insight,
a key to the complete understanding of the slaving small-scale dynamics will be found in the physical space structure of the (covariant) Lyapunov vectors~\cite{KO,IKTY,ITY, Ginelli,lucarini} corresponding to the Lyapunov exponents;
these are `unstable modes' of turbulent structures, such as the hierarchy of antiparallel vortex tubes~\cite{GSK2017,YGT2022},
as will be presented elsewhere.
These future studies based on the proposed framework can lead to new DA algorithms, including an approach for stabilizing the unstable direction of the DA manifold.

In the rapid development phase of the data-driven methods for turbulence~\cite{MIG, Caulfield, Bruntonbook}, including the DA methods~\cite{Titi, Li2020, YYK2005,Lalescu2013, Martin2021, Zaki, Biferale},
the dynamical system approaches are significant.
In particular, a neural network-based study of turbulence modeling found a qualitative change in modeling difficulty at $k^{\ast}=0.2/\eta$~\cite{MIG}; the modeling of turbulent dynamics in the wavenumber region higher than that is feasible without difficulty, which can be understood within the proposed framework using TLEs.
The TLEs will provide insights into the data-driven science of turbulence and, more generally, high-dimensional chaotic dynamical systems with hierarchical spatiotemporal scales.

This work was partially supported by JSPS Grants-in-Aid for Scientific Research (Grants Nos.~22K03420, 22H05198, 20K20973, 20H02068, 19K14591, and 19KK0067).
Direct numerical simulations of the Navier--Stokes equations were conducted using supercomputer systems of the Japan Aerospace Exploration Agency (JAXA-JSS2).


\nocite{*}


\clearpage

\end{document}